# A REVIEW OF EFFECTS OF CLIMATE CHANGE ON AGRICULTURE IN AFRICA

Samuel Asante Gyamerah[a,b], Dennis Ikpe[c,d]

[a]*Department of Statistics and Actuarial Science, Kwame Nkrumah University of Science and Technology, Kumasi-Ghana*

[b]*Laboratory for Interdisciplinary Statistical Analysis – Kwame Nkrumah University of Science and Technology (KNUST-LISA), Kumasi-Ghana*

[c]*Department of Probability and Statistics, Michigan State University, Michigan, USA*

[d]*Africa Institute for Mathematical Sciences, South Africa*

Agriculture plays critical roles by serving as the fundamental source of food and nutrition, livelihoods, and vehicle for growth and development to many countries around the globe. The agriculture sector forms the backbone of economies of many low to middle income countries especially in Africa. The sector's contribution to GDP varies across African countries, however, average contribution is 15% according to OECD-FAO (2016). Tongwane & Moeletsi, (2018) suggests a staggering 32% contribution to GDP of Africa by agriculture alone. The significant economic contribution could partly be owed to the less diverse nature (Osakwe, 2007) of most African economies. In sub-Saharan Africa, the sector employs 65% of the labor force of the continent in addition to the invaluable contribution it makes to food security and poverty reduction (Tongwane & Moeletsi, 2018; Chuku & Okoye, 2009). Improvement in the sector is therefore imperative (Bachewe, Berhane, Minten & Taffesse, 2018) to achieving food security, significant reduction in poverty and propelling the engine of growth. This places the agriculture sector at the center-stage for meeting both regional and global goals such as the Agenda 2063 and the Sustainable Development Goals (SDGs) in the continent. Transformation of the sector is therefore a dire need to be pursued as part of the broader economic transformation required to sustain the continents rapidly growing population.

Until recently, the agricultural systems in many African countries were largely characterized by moribund practices. Most agricultural activities in the continent are carried out at the subsistence level. According to Sibhatu & Qaim (2017), smallholder farmers contribute largely to the quantity of the food produced in

Correspondence: saasgyam@gmail.com

Africa and Asia. Despite their enormous contribution to food production, their reliance on traditional methods, which increasingly become obsolete stifles productivity in the agriculture sector (Mwangi & Kariuki, 2015). Nevertheless, commercialization and general transformation of the sector in Africa has been significant over the past decade despite less notice (Jayne & Ameyaw, 2016). Studies (Chauvin, Mulangu & Porto, 2012; Salami, Kamara & Brixiova, 2010; Fuglie & Rada, 2013) blame the slow transformation of the sector according to Tongwane & Moeletsi (2018) several factors including but not limited to low technology adoption; challenges related to market access; less skilled labor and poor soil fertility. (Generally, productivity is in a steady increase as medium-sized to large-sized farms begin to increase in rural areas. Unfortunately, this is threatened by several factors including declining soil fertility, increasing cost of arable land, and the impacts of climate variability and change (Jayne & Ameyaw, 2016). The vulnerability of the continent's slowly transitioning agricultural system (Chauvin, Mulangu & Porto, 2012) to projected impacts of climate change and variability is indubitable as the system relies on climate—rainfall, light, temperature, humidity other climatic variables (Molua, 2002) for food production. Further, the agriculture system is increasingly becoming sensitive to adverse impacts of climate change and variability due to lack of modern technologies and low adoption climate-smart agricultural methods. These dynamics opens up the very backbone of every African economy, and mainstay to many vulnerable rural populations to dire consequences of projected climate impacts.

Currently, agriculture in Africa contributes only a tenth to global Green House Gas (GHG) emissions from agriculture (Tongwane & Moeletsi, 2018). Despite its relatively low contribution to GHG, a conundrum of "climate justice", adverse impacts of climate change disproportionately threaten Africa's agriculture, the continent's main economic sector. Rapidly changing rainfall patterns resulting in high unpredictability, is one of the major threats to agricultural systems in the continent. Famers lack the requisite technology, infrastructure, and knowledge and skills among others to effectively cope or adapt to drastic short-term changes in weather patterns known as "climate variability", and seasonal variations which are increasingly becoming observational. Aside from their low adaptive capacity, other vulnerabilities such as the characteristic rain-fed agriculture dominance in SSA (Kotir, 2011) and many other African countries exposes them more to climate risk. Climate variability potentially induces extreme conditions such as water stress and erosion among others which threaten plant growth and health, consequently dwarfing crop production. According to Müller, Cramer, Hare & Lotze-Campen (2011), the IPCC Synthesis Report (SYR) and the Summary for Policy Makers of the Working Group II Report indicated that climate change and variability could result in close to 50% decline in rain-fed agricultural crop production by 2020. This underscores the need for African countries aggressively pursue swifts to irrigation and climate-smart

Correspondence: saasgyam@gmail.com

agriculture systems in order to sustain the sector and prevent knock-on effects food security, livelihoods, economic growth and development.

Aside from the effects from the increasingly variable and unpredictable climate on agriculture, the impacts of long-term climate events are enormous. Extreme climate change-induced phenomena such as increased intensity and frequency of drought and floods, and ENSO (El-Nino) which usually occur as a result of long-term changes pose severe challenges to agriculture in Africa. According to Sylla, Nikiema, Gibba, Kebe & Klutse (2016), the current GHG emission scenarios would result in "shorter rainy seasons, generalized torrid, arid and semi-arid conditions, longer dry spells and more intense extreme precipitations" in most West African countries. Climate change increases not only the vulnerability of the agriculture system in Africa drought and floods but also their impacts. These events, caused by extreme changes in temperature and rainfall are already observational many regions of the African continent (Ngaina & Mutai, 2013; Otto et al., 2015; Kadomura, 2005). Crop, fisheries and animal production are among the major components of agriculture in Africa that are highly sensitive, and have suffered (Blanc, 2012; Brida, Owiyo & Sokona, 2013; Zougmoré et al., 2016) from the impacts of these events, especially, drought and floods.

Further, according to Boko et al. (2007) the overall revenue Africa makes from crop production is projected to decline by up to 90% by 2100 due to climate change (Williams & Kniveton, 2011). This would be enhanced in part by substantial reduction in the length of cropping seasons and unpredictability of initiation of wet periods (Williams & Kniveton, 2011). Drastic seasonal shifts of this nature would drive declines in crop production as a result of potential end of rainfall season when crops are yet to mature or torrential rainfalls during periods of crop harvest or less requirement for excess moisture. Whiles improved seeds with the capacity to adapt to such changes and precision agriculture are touted around the world, most African countries unfortunately lags behind in the adoption of biotechnology and advanced engineering among other critical technologies (Muzari, Gatsi & Muvhunzi, 2012). Also, a majority of farmers in the continent are less skilled, and mainly rely on traditional farming methods and implements. Moreover, the impacts abrupt seasonal changes are not limited to reduction in crop production, animal or livestock production could equally be hampered as adverse climate change impacts could affect incidence of disease and natural resources on which livestock depend on (Baumgard et al., 2012). Also, the fisheries sector is not immune as upwelling regimes among other important phenomena could be significantly altered by such climate change induced variations (Bakun etal., 2015). Access to credit remains challenging to these vulnerable famers as investment in the sector fails to meeting growing demand for expansion. These factors together complicate the adaptive capacity of farmers, especially, small-holder farmers who constitute a significant proportion, against increasing uncertainties related to season and climate variables. Finally, the

Correspondence: saasgyam@gmail.com

Fifth Assessment Report of the Intergovernmental Panel on Climate Change (IPCC), African economies relies on climate-sensitive sectors, especially agriculture, fisheries and forestry climate change which could suffer multiplied impacts of existing stressors and emerging adverse conditions hence the need for integration of adaptation planning in decision-making (Carabine, Lemma & Dupar, (2014). This would not only be instrumental in building resilience but also contribute significantly to the reduction in hunger and poverty; and accelerating growth and development all of which are pivotal to the global agenda for sustainable development—Sustainable Development Goals.

Correspondence: saasgyam@gmail.com

Correspondence: saasgyam@gmail.com